\documentclass[12pt, a4paper, oneside]{article}
\usepackage{graphicx}
\usepackage{color}
\usepackage[margin=1.0in]{geometry}

\begin{document}
\thispagestyle{empty}
\title{Visualization of short-term heart period variability with network tools as a method for quantifying autonomic drive}
\author{ Danuta Makowiec$^{ a,*}$, 
         Beata Graff$^b $, 
         Agnieszka Kaczkowska$^c $, \\
         Grzegorz Graff$^c $,
         Dorota Wejer$^a $, 
         Joanna Wdowczyk$^d $, \\
         Marta \.Zarczy\'nska-Buchowiecka$^d $, 
         Marcin Grucha\l a $^d $, 
         \bigskip Zbigniew R. Struzik$^{e,a} $ \\
            \small $^a$Institute of Theoretical Physics and Astrophysics \\
            \small University of Gda\'nsk\\ \bigskip
            \small 80-952 Gda\'nsk, ul. Wita Stwosza 57, Poland\\
            \small  $^b $ Hypertension Unit, Department of Hypertension and Diabetology \\
            \small  Medical University of Gda\'nsk\\    \bigskip
            \small 80-952 Gda\'nsk, ul. Debinki 7c, Poland\\
            \small $^c$Faculty of Applied Physics and  Mathematics \\
            \small Gda\'nsk University of Technology\\  \bigskip
            \small 80-233 Gda\'nsk, ul. Narutowicza 11\slash 12, Poland\\
            \small $^d$1st Chair \& Clinic of Cardiology \\
            \small Medical University of Gda\'nsk\\ \bigskip
            \small  80-211 Gda\'nsk, ul. Debinki 7, Poland\\
            \small $^e$ RIKEN Brain Science Institute, Japan  \\   \bigskip
            \small Graduate School of Education, The University of Tokyo, Japan\\ 
            \small $^*$ corresponding author
          }
\maketitle

\subsection*{Acknowledgements}
The authors DM, ZRS, BG, DW and AK acknowledge the financial support of the National Science Centre, Poland, UMO: 2012\slash 06\slash M\slash ST2\slash 00480.\\

This work was partially realized under the SKILLS project of the
Foundation for Polish Science (172/UD/SKILLS/2012) and was co-financed by the European Union through the European Social Fund.

\newpage
\subsection*{Abstract}
{\bf Introduction and purpose}\\
Signals from  heart transplant recipients can be considered to be a natural source of information for a better understanding of the impact of the autonomic nervous system on the complexity of heart rate variability. Beat-to-beat heart rate variability can be represented as a network of increments between subsequent $RR$-intervals, which makes possible the visualization of short-term heart period fluctuations.

{\bf Methods}\\
A network is constructed of vertices representing increments between subsequent $RR$-intervals, and edges which connect adjacent $RR$-increments. Two modes of visualization of such a network are proposed.  The method described is applied to nocturnal Holter signals recorded  from healthy young people and from cardiac transplant recipients.  Additionally, the  analysis is  performed on surrogate data: shuffled RR-intervals (to display short-range dependence), and shuffled phases of the Fourier Transform of RR-intervals (to filter out linear dependences).   

{\bf Results}\\
Important nonlinear properties of autonomic nocturnal regulation in short-term variability in healthy young persons are associated with $RR$-increments: accelerations and decelerations of a size greater than about 35 ms. They reveal that large accelerations are more likely antipersistent, while large decelerations are more likely persistent. 

Changes in $RR$-increments in a heart deprived of autonomic supervision are much lower than in a healthy individual, and appear to be maintained around a  homeostatic state, but there are indications that this dynamics is nonlinear.

{\bf Conclusion} \\
The method is fruitful in the evaluation of the vagal activity - the quantity and quality of the vagal tone - during the nocturnal rest of healthy young people. The method also successfully extracts nonlinear effects related to intrinsic mechanisms of the heart regulation.

Since the vagal part of autonomic regulation is considered to be responsible for nocturnal heart rhythm with large accelerations and decelerations, we hypothesize that vagal activity is a crucial source of the complexity in short-term heart rate variability.

\newpage
\section{Introduction}
Network methods have been successfully used to capture and represent properties of multilevel complex  man-made systems \cite{Havlin.2012} and living organisms \cite{Bashan.2012}.  The use of network representations in the characterization of time series complexity is a relatively new but quickly developing branch of time series analysis (\cite{Donner.2010}, \cite{Fortunato.2010}). The most direct method is to map a time series into a graph in which vertices represent signal values, while edges link values that are consecutive in a signal. The correspondence between time series formed by consecutive cardiac interbeat intervals, so called $RR$-intervals, and such networks  was studied by Campanharo et al. \cite{Campanharo.2011}. The topology in these networks appeared as a clique, i.e., each state is reachable from each other in a single step. Understanding $RR$-interval dynamics arising from a network with  such a structure is not straightforward. It appears,  in general, that information provided by a network graph strongly depends on the nature of sequences and our knowledge about the underlying dynamics (\cite{Fortunato.2010}, \cite{Havlin.2012}). Therefore the use of network methods, e.g., visualization or/and structure decomposition, is effective only if they are used in conjunction with other sources of learning. We show how tools developed within the scope of complex networks can be fruitfully applied to the qualification and quantification short-term heart period dynamics. 

Fluctuations in $RR$-intervals are  known to have a scale-invariant structure which demonstrates fractal (\cite{Kobayashi.1982}, \cite{Yamamoto.1991}, \cite{Peng.1995}) and multifractal (\cite{Ivanov.1999}) properties. These fluctuations appear as a result of many component interactions acting over a wide range of time and space scale. Competing stimuli from the autonomic nervous system are assumed to be the reason for the fractal organization observed in $RR$-intervals \cite{Struzik.2004}. By observing subsequent changes in $RR$-intervals, we obtain  beat-to-beat information about the resulting force of these interactions. We have found that important  dynamical aspects about the autonomic competitive regulation  can be  described   by changes in $RR$-intervals, namely by $RR$-increments (\cite{Makowiec.2013}, \cite{Makowiec.2014}).

Signal increments can be decomposed into their magnitude (absolute value) and their direction (sign). Magnitude and sign analysis has been used to investigate the scaling properties of RR-intervals \cite{Ashkenazy.2001}. It has been found that magnitude series are long-range correlated, while sign series are anticorrelated. Furthermore, it has also been shown that during sleep, the strength of these correlations varies depending on the stage of the sleep: rapid-eye-movement (REM) or other (non-REM) sleep stages \cite{Kantelhardt.2002}. It appears that both the strongest anticorrelations in the sign signals, and largest exponents for long-range correlations for the magnitude signals are in REM sleep. Furthermore, it is assumed that during REM sleep, the nonlinear properties of the heartbeat dynamics are more pronounced.

During sleep, the heart rate is mostly regulated by the autonomic nervous system and is less influenced by physical or mental activity. Moreover, during the night, vagal (parasympathetic) predominance is present, which makes this period a useful state to observe autonomic activity. The classical tools applied to measuring heart rate variability during physiological sleep have shown that the REM stage is characterized by a likely sympathetic predominance associated with a vagal withdrawal, while the opposite trend is observed during non-REM sleep (see \cite{Tobaldini.2013} for a review). Previous studies have also shown that alternations in nocturnal heart rate variability  have clinical importance, e.g., may explain why sudden cardiac death in many cases occurs during the night \cite{Vanoli.1995}. 

Heart transplantation  surgery destroys the nerve connections between the organism and the graft --- the donor heart is completely denervated, the vagal ganglia at the sinus node are cut off from medulla signals. The regulation is driven by the intrinsic heart mechanisms. The sympathetic control works indirectly at the level of circulating adrenergic hormones in the blood.  The lack of vagal activity has the effect, for example, that heart transplant recipients have a resting heart rate higher than the average in healthy people, and their heart rate variability is  significantly reduced \cite{Bigger.1996}. The exception is a small respiratory sinus arrhythmia (\cite{Radealli.1996}, \cite{Eckberg.2003}), which is assumed to be an effect of the intracardiac reflex (\cite{Armour.2008}, \cite{Zarzoso.2013}) or mechanical stretch of the sinus node. Cardiac reinnervation has been demonstrated in long-term heart transplant recipients (\cite{DeBorne.2001}, \cite{Porta.2011}, \cite{Cornelissen.2012}), but it seems that it is limited to the sympathetic nerves. Therefore, the comparison of the nocturnal heart rate variability in healthy young individuals and heart transplant patients gives a unique opportunity to show the impact of autonomic (especially vagal) activity on heart rate regulation.

Following \cite{Ashkenazy.2001}, to discover in which way properties of networks constructed from $RR$-increments demonstrate nonlinear or/and linear dependences among consecutive $RR$-intervals, we investigate properties of artificially modified  $RR$-interval data \cite{Schreiber.2000}. In the following, we argue that network methods are successful in detecting nonlinear properties in the dynamics of autonomic nocturnal regulation in short-term variability. 
Two modes of visualization of networks constructed from $RR$-increments are proposed. The first is based on the handling of a state space. The state space of  $RR$-increments can be modified by a bin size used to code a signal, and by the role of a given vertex  as the representation  of events  occurring in a signal. The second mode relies on the matrix representation of the network on the two-dimensional plane. This approach is similar to the accepted method, known as the Poincar\'{e} Plot representation of time series for evaluation of heart rate variability. The methods introduced will be applied to nocturnal Holter signals recorded  from healthy young people and  from  cardiac transplant recipients. Thus we obtain a way to filter out the intrinsic heart variability from the autonomic drive and then to quantify complexity in the short-term $RR$-interval variability related to nocturnal rest. Changes in $RR$-increments in a heart deprived of autonomic supervision provide us with insight into beat-to-beat dependences in forces governing the intrinsic heart dynamics. 

\section{Method}
\subsection{Groups and signals studied}
Twenty-four-hour Holter ECG recordings during a normal sleep-wake rhythm were analyzed in two study groups. The first group, the {\it Young}, consisted of healthy young volunteers (18 female, 18 male, age 19-32). The second group, the {\it HTX}, comprised heart transplant  patients (surgery at ages 28 to 65).  Data from the {\it HTX} group was constructed of 20 recordings obtained from 10 patients without any signs of heart graft rejection and who had undergone surgery more than 12 months previously. The Holter recordings  were  first  analyzed using Del Mar Reynolds Impresario software and screened for premature, supraventricular and ventricular beats, missed beats and pauses. Finally, the signals were thoroughly manually corrected and annotated.

Since the analysis concentrates on hours of sleep, the $RR$-intervals were analyzed from 24:00 to 04:00 in the case of the $Young$ group and from 22:00 to 05:00 in the case of  signals from the $HTX$ group.

As the method of signal preprocessing may impact the results, only long, good-quality fragments of ECG were analyzed. Namely, the parts  which contained more than five hundred normal-to-normal $RR$-intervals were extracted and then joined together in a sequence of 10000 $RR$-intervals.   What is worth noting is that all the signals constructed were built from less than seven consistent parts. The number of 10000 points was chosen to ensure proper statistical relevance.

\subsection{Signal preprocessing}
Our Holter equipment provides data with a 128 Hz sampling frequency. Therefore, the $RR$-intervals have a limited resolution: 1s/128=7.8125 ms, which can be approximated as $8$ ms. This value, denoted as $\Delta_0$,  is  accepted as  the signal resolution.  To decrease the number of different values appearing in a sequence of $RR$-intervals, we use a binning procedure based on multiples of $\Delta_0$. Namely, we always set the bin size to $\Delta_{bin}$ as  $\Delta_{bin}= k\Delta_0$ for $k=1,2,\dots$.  The bin quantization described has the effect that $RR$-intervals take values which are multiples of the bin size $\Delta_{bin}$. As a consequence,  $RR$-increments are also multiples of $\Delta_{bin}$, namely, $\Delta RR_t\in\{0, \pm \Delta_{bin}, \pm 2\Delta_{bin}, \pm 3\Delta_{bin}, \dots\}$. 

The two types of  artificially modified cardiac signals were constructed for their further use in statistical tests:
\begin{itemize}
    \item[] {\it shuffled} signals, which  were obtained by random shuffling of $RR$-intervals;
    \item[] {\it surrogate} signals, which were calculated by randomization of phases in the Fourier transform of $RR$-intervals.
\end{itemize}

The analysis of {\it shuffled} signals tests the presence of dependencies in the signals studied, while the analysis of {\it surrogate} signals  provides information about whether these dependences are linear or not \cite{Schreiber.2000}.  Signals of both types were prepared with the help of the TISEAN software \cite{TISEAN}.
For each cardiac signal, we prepared ten shuffled and ten surrogate signals.

A network was constructed separately for each signal analyzed. Then the mean network for each of the groups of signals studied was established by collecting networks corresponding to the same class of subjects. The confidence interval (CI) for each element of  the mean network was also estimated. Calculations were performed with the special software prepared by us. 

\subsection{Transition network  for  $RR$-increments}
Let ${\bf RR_{\Delta_{bin}}}=\{RR_0, RR_1, \dots,RR_t,\dots, RR_N\} $ be a time sequence of $RR$-intervals binned in a $\Delta_{bin}$. Let ${\bf \Delta RR}=\{\Delta RR_1,\Delta RR_2, \dots,\Delta RR_N\} $ be a time sequence of   $RR$-increments, i.e., $\Delta RR_t= RR_t-RR_{t-1}$. Discrete values of the set $\bf \Delta RR$ serve as states in the state space of the transition network indexed by the bin value $\Delta_{bin}$. 

Let $K$ denote the number of different states in the network state space, and let us arrange them as follows. If the smallest state  is  $\Delta^{min}= \min_t \bf \Delta RR$ and   the greatest state is $\Delta^{max}=\max_t \bf \Delta RR$, then the vertices of a network are labeled consequently as:
\begin{equation}
\begin{array}{rcl}
 \Delta^{(1)} =\Delta^{min},\quad \Delta^{(2)}=\Delta^{(1)}+\Delta_{bin} ,\quad\dots, \quad
  \Delta^{(K)}=\Delta^{max}= \Delta^{(1)}+(K-1)\Delta_{bin} .
  \label{order}
\end{array}
 \label{RR-labels}
\end{equation}
A directed edge $(\Delta^{(I)},\Delta^{(J)})$ from a  vertex  $\Delta^{(I)}$ to a vertex $\Delta^{(J)}$ is established if $\Delta^{(I)}$ and $\Delta^{(J)}$ represent a pair of  consecutive events in a time sequence $\bf\Delta RR$. Namely, there is a moment in time $t=1,\dots N-1$,  for which $(\Delta RR_t , \Delta RR_{t+1})= (\Delta^{(I)}, \Delta^{(J)})$. If a given pair of increments occurs many times in $\bf \Delta RR$, the weight of this edge $ w(\Delta^{(I)}, \Delta^{(J)})$  increases accordingly to represent counts of occurrences.

Note that  the weight of the edge $ w(\Delta^{(I)}, \Delta^{(J)})$ measures the size of a set consisting of the following events:
\begin{eqnarray}
\label{three-events}
w(\Delta^{(I)}, \Delta^{(J)})&=& |\{ (RR_{t-1}, RR_t, RR_{t+1})  
\colon  \mbox{ where} \\ \nonumber
\Delta^{(I)}= RR_t - RR_{t-1},&& \quad  \Delta^{(J)}= RR_{t+1} - RR_{t}
\quad \mbox{ for } t= 1,\dots N-1 \} |.
\end{eqnarray}

This means that:
\begin{itemize}
\item[] if $ \Delta^{(I)} \cdot \Delta^{(J)} > 0$, hence  both increments are negative  or both are positive, we observe a run of accelerations or decelerations, accordingly;
\item[] if $ \Delta^{(I)} \cdot \Delta^{(J)} < 0$, we observe an alternation between an  acceleration and a  deceleration or vice versa.
\end{itemize}

This completes the construction of the transition network from a given time series. The resulting network  is  directed and weighted. The sums of weights of edges adjacent to a given vertex  (total number of  $incoming$ and total number of $outgoing$ edges) provide the  basic  network characteristics (called $in$- degree and $out$- degree, respectively), which quantify the role of the vertex in a network. But a network constructed from time series is specific in that each $outgoing$ edge from a given vertex is accompanied  by an edge $incoming$ to this vertex (with the exception of vertices representing the first and last events in consistent parts of a signal), which implies that the $in$ and $out$ degrees of each vertex can be considered to be equal to each other. This degree, if normalized by the length of time series, is directly related to the probability $p$ that an event represented by $\Delta^{(I)}$ occurs in a signal.

The modular structures, also called the community structure, in networks have been shown to be relevant to the understanding of the structure and dynamics of the system studied \cite{Havlin.2012}. However, this problem has been found to be difficult and has not yet been satisfactorily solved (\cite{Kumpula.2008}, \cite{Fortunato.2010}). Here we propose to investigate modularity in the transition network by the so-called {\it $ p$-core graph} (\cite{Seidman.1983}, \cite{Kumpula.2008}). The $p$-core graph is constructed from a given network by the removal of all the vertices with a probability less than $p$. Then all the edges which connected these deleted vertices with the other parts of a network are removed.  The sum of  normalized weights in the resulting subgraph is called the {\it volume of the $p$-core graph}.  A decay in this volume with an increasing $p$ value is known as the network disintegration \cite{Makowiec.2013}.

\subsection{Transition network graph}
\label{sec:Transition}
Visualization of a transition network is challenging because usually a transition network consists of many vertices which are densely, often completely, interconnected. The plot of such a network may be barely readable. Therefore the graph organization requires a special effort.

There are parameters in the method which have to be thoroughly tuned:
\begin{itemize}
\item[]$\Delta_{bin}$ --- the bin size which is used in preprocessing RR-intervals and which determines the number of states in the state space;
\item[]$p$ --- the probability of neglected events, which also allows a reduction in the number of states.
\end{itemize}

Since states in  the state space are ordered according to the values of their labels, see Eq. \ref{order}, we  plot them in a circle arranged clockwise according to increasing value of the vertex label  from $\Delta^{min}$ to $\Delta^{max}$. Moreover, if we call two vertices $\Delta$-neighboring when the magnitude of difference between their labels is equal to $\Delta$, then we can code transitions between $\Delta$-neighboring vertices by colors.

Here we  use the following color code:
\begin{itemize}
\item[-]    violet to mark $0$-neighboring vertices, i.e., loops describing events  of two adjacent accelerations or decelerations of the same value; the case of the $0\rightarrow 0$ loop denotes the situation when three consecutive $RR$-intervals have  the same value;
\item[-]    green  to mark $\Delta_{bin}$-neighboring vertices, i.e., transitions to the nearest neighbors in the state space; they denote the smallest possible observable changes in subsequent accelerations and/or decelerations  within a given binning;
\item[-]  blue to mark transitions  for $2\Delta_{bin}$-neighboring vertices;
\item[-] red to mark  transitions between  $3\Delta_{bin}$-neighboring vertices;
\item[-] yellow to mark the transitions linking  $4\Delta_{bin}$-neighboring vertices;
\item[-] black to mark transitions of a size larger than $4\Delta_{bin}$ which, for example, in the case of $\Delta_{bin}=8$ ms means changes of at least of 40 ms.
\end{itemize}
Moreover, we use also the width of an edge to visualize the weight of a given transition.

In the following, we use the popular software PAJEK \cite{PAJEK} to plot graphs of transition networks.

\subsection{Matrix representation of a transition network}
Adjacency matrices and transition matrices are standard representations of any network \cite{Fortunato.2010}. For a transition network with $K$ vertices, the adjacency matrix $\bf A$ is a $K \times K$ matrix. The number of the outgoing edges from vertex $\Delta^{(I)}$ to vertex $\Delta^{(J)}$ is counted and designated as $A_{(I)(J)}$. If there is no edge between these vertices, then $A_{(I)(J)}=0$. Hence
\[
A_{(I)(J)}=\left\{
\begin{array}{ll}
w(\Delta^{(I)}, \Delta^{(J)}) &\textrm{total number of  edges from}\; \Delta^{(I)}\; \textrm{to}\; \Delta^{(J)};\\
0 & \textrm{in other cases}.\\
\end{array}
\right.
\]
In the following, we normalize counts $ w(\Delta^{(I)}, \Delta^{(J)}) $ by the total number of events. As a results,  $A_{(I)(J)} $ means the probability of a given transition. Referring to a signal with $RR$-increments $A_{(I)(J)} $  stands for the probability that value $\Delta^{(J)}$ occurs after $\Delta^{(I)}$ in a signal.
The transition matrix $\bf T$ is obtained by dividing elements of  each row $(I)\colon(1),\dots, (K)$ of the  matrix $\bf A$   by  the total weight  of  vertex $\Delta^{(I)}$. Thus,
$$T_{(I)(J)}= \frac{ w(\Delta^{(I)}, \Delta^{(J)})} { \sum_{\Delta^{(I)}} w(\Delta^{(I)}, \Delta^{(J)})}. $$
Therefore $\bf T$ describes a Markov walk on a network where a walker being in   vertex $\Delta^{(I)}$ moves to $\Delta^{(J)}$ with a probability $T_{(I)(J)}$.

It appears that the contour plots of  adjacency and transition matrices  provide a readable visualization of transition networks obtained from  $RR$-increments  even in the case when the bin is  equal to the resolution  of signals. Manipulations in the range of the axes allow one to pass through the whole range of values obtained. However, when departing from  $(0,0)$, less probable events with larger standard errors are estimated. The large variations between neighboring points lead to an unclear picture if the plots are constructed from signals binned in a small bin size.  Therefore, in the following, we limit our interest to the ranges of $RR$-increments  which contain the most probable events.

\begin{figure}[h!]
\centering
   \includegraphics[width=0.9\textwidth ]{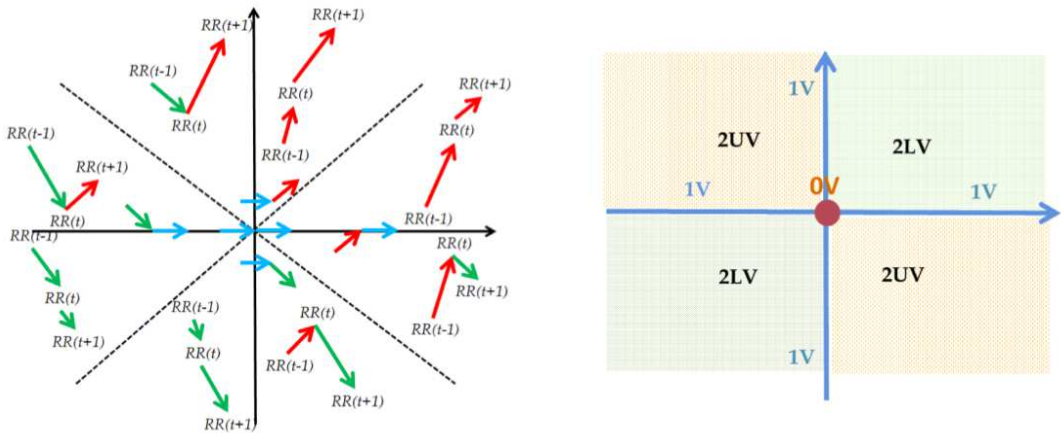}
   \caption{The patterns of changes in $RR$-intervals corresponding to particular parts of the matrix representation of a network of $RR$-increments (left) and their interpretation as  variations 0V, 1V, 2LV and 2UV --- codes proposed by Porta et al. \cite{Porta.2007} (right). Red arrows indicate decelerations, green arrows denote accelerations, blue arrows correspond to no-change events.}
   \label{fig:patterns}
\end{figure}

Each point  $(\Delta^{(I)}, \Delta^{(J)})$ of the contour plots can be resolved into the three $RR$-interval patterns as described by formula (\ref{three-events}). Moreover, these  events can be translated into  codes of short-term variability proposed by Porta et al. \cite{Porta.2007}: $0V$ - 0 variation, $1V$ - 1 variation,  $2LV$ - 2 likely variations and $2UV$ - 2 unlikely variations. The relation between three $RR$-interval patterns and their description by 0, 1 or 2 variations is shown in Fig. \ref{fig:patterns}.

\section{Results}
\subsection{Graphs of transition networks for the {\it Young}}
In the presentation of our results, we first refer to some graphs which demonstrate the possibilities of the method of visualization introduced. In particular, these graphs clarify the influence of the parameters $\Delta_{bin}$ and $p$ on the graph shape. 

\begin{figure}[h!]
\centering
   \includegraphics[width=0.9\textwidth ]{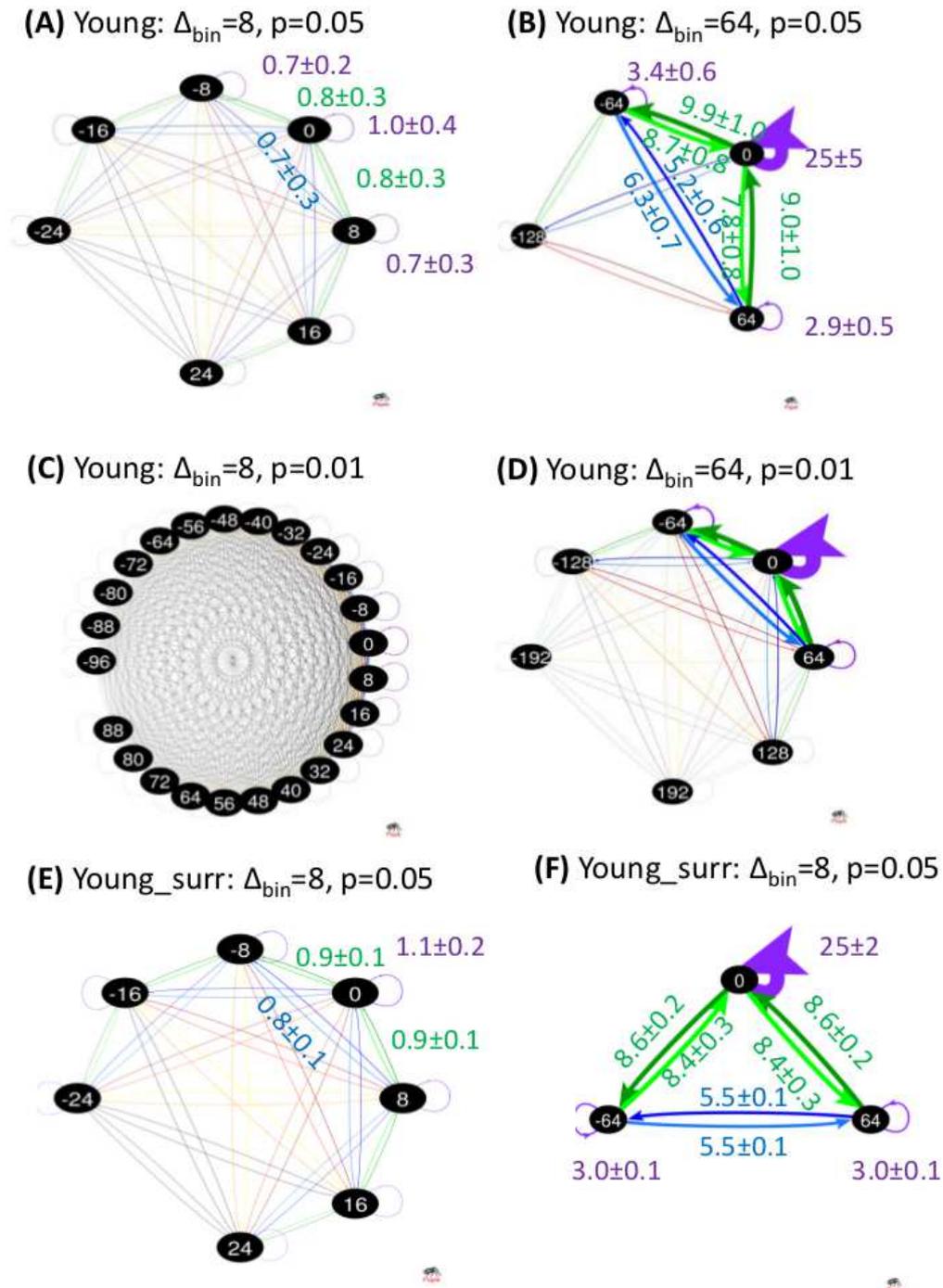}
    \caption{The $p$-core graphs  with  $p=5\%$ (A, B, E, F) and $p=1\%$ (C,D) of the mean networks obtained from signals of young persons binned in $\Delta_{bin}=\Delta_0=8$ ms  (A,C) and $\Delta_{bin}=8\Delta_0=64$ ms (B,D), and surrogates obtained from signals from young persons (E,F). The crucial transitions - the thickest edges - are described by the mean of probability given in per cents $\pm$ 95\% CI. (C and D have the same weights as A and B, respectively). The colors of the values match the colors of the edges.}
    \label{fig:young}
\end{figure}

In Fig. \ref{fig:young}, there are six graphs which represent networks obtained from signals of the {\it Young} group (A)-(D) and their surrogates (E), (F). The left column networks were prepared with signals binned in $\Delta_{bin}=8$ ms, while the right column graphs come from signals binned in $\Delta_{bin}=64$ ms. The first and third row graphs show vertices which correspond to deceleration/acceleration events appearing in a signal with probability  $p > 5\%$. The second row graphs contain vertices representing more rare events  because only vertices with  $p>1\%$ are plotted. 

All the graphs are complete, which means that each vertex is connected to all others. The color of the edges corresponds to size of the change in a way defined in Sec. \ref{sec:Transition}. The weight of an edge is represented by their width.  The global weights of the most important transitions are   described additionally by giving their probability value in per cents and $\pm 95\% $ CI. 

We see that the $(0, 0)$ transition is the most frequent in all the graphs, namely the transition from no-change to no-change occurs the most often. However the probability of observing such an event is different depending on the binning applied to the signals. The bin size works like a magnifying glass, making it possible to perceive a single event in greater detail. For example, vertex $0$ in the network constructed from signals binned at 64 ms (B) represents the whole $p$-core graph (A) obtained at $p=5\%$. 

Fig. \ref{fig:young} also gives graphs obtained from networks constructed from  surrogates of cardiac signals of the {\it Young} group. They look similar to graphs obtained from original signals, which might indicate that dynamical relationships among the most important events are of a linear type. We do not present graphs obtained for the shuffled signals of the {\it Young} group because they are barely readable.

Moreover, when signals are binned in $\Delta_0$, the $p$-core graph of  $p=5\%$ consists of only a few vertices (A), while the number of vertices  grows sharply if we decrease the probability $p$ to $1\% $ (C). Hence, graph (A) rather superficially describes the dynamics of the system, since a slight change in any parameter strongly influences the graph shape. This is different  in the case of signals binned in $\Delta_{bin}= 64$ ms; compare (B) to (D). Therefore the volume of a given  $p$-core graph gives us an important message about how dynamical forces presented by a graph are meaningful  for the overall dynamics of the system studied. Here, the volumes of the $p$-cores presented in Fig. \ref{fig:young} are (A): $ 27\pm 14\% $,  (B): $ 89.1\pm 1.0\% $, (C): $ 84.7\pm 1.4\% $, (D): $ 97.1\pm 0.2\% $, (E): $ 26.3\pm 3.2\% $,  (F): $ 76.2\pm 1.0\% $ (mean $\pm 95\% CI$).

\subsection{Graphs of transition networks for {\it HTX}}
\begin{figure}[htbp]
    \includegraphics[width=1\textwidth ]{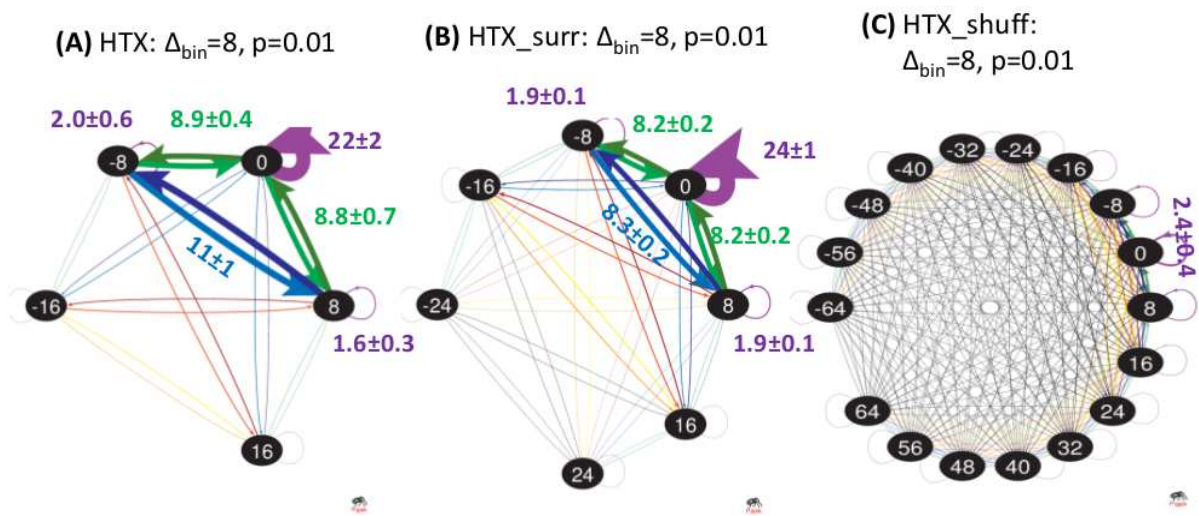}
    \caption{The $p$-core subgraphs of the mean transition networks  obtained from signals from {\it HTX} (A) and their surrogates (B) and shuffled (C) at $p=0.01$ in bin $\Delta_0=8 ms$. The most important transitions - the thickest edges - are described by their probabilities given in per cents $\pm$ 95\% CI.}
\label{fig:htx}
\end{figure}

The signals with $RR$-intervals obtained from patients after $HTX$ are plain in the sense that consecutive $RR$-increments do not differ much. Therefore, it becomes possible to present readable $p$-core graphs even when  $p$ is low, e.g., $p=1\%$,  and the bin size is equal to the signal resolution $\Delta_{bin}=8$ ms. Moreover,  graphs obtained from modified  {\it HTX} signals, their  surrogates and when shuffled, are also clear. All these graphs are shown in Fig. \ref{fig:htx}.  

Figure \ref{fig:htx} demonstrates the following:
\begin{itemize}
\item How different the graph obtained from {\it HTX\_shuffled} is from all the other graphs. This indicates that  adjacent $RR$-intervals resulting from  intrinsic mechanisms of variability are strongly correlated;
\item There is similarity between graphs representing signals {\it HTX} and {\it HTX\_surrogates}. However we should note that there is a noticeable difference in the probability of transitions between no-change vertex $0$ and vertices $0, \pm 8$.  This observation may indicate that  dependences  between  $RR$-intervals cannot be approximated by some linear relations;
\item There is a similarity between graphs of {\it HTX} and those of the {\it Young} which are binned at an interval 8 times longer than the {\it HTX} series; see Fig. \ref{fig:young}(D). This approximate similarity may give rise to the conjecture that the variability driven by autonomic regulation enhances eight times the magnitude of fluctuations of the intrinsic heart period variability. 
\end{itemize}

\subsection{Network disintegration }
\begin{figure}[htbp]
    \centering
    \includegraphics[width=0.75\textwidth ]{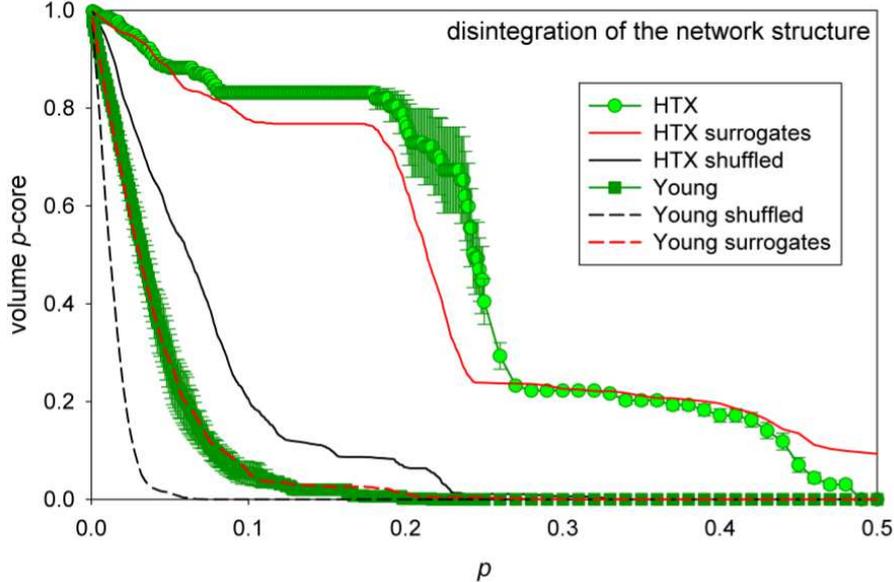}
    \caption{Network disintegration by the volume of $p$-core resulting from  subsequent removal of vertices with probability less than $p$. }
    \label{fig:disintegration}
\end{figure}

The decay of the volume of the $p$-core when $p$ is increasing can provide us information about the collective  structures in the system dynamics. In Fig. \ref{fig:disintegration}, we show these decays for all signals studied binned in $\Delta_{bin} =8$ ms.

It appears that the network formed from  signals of the {\it HTX } group is significantly more resistant to the vertex removal than all the other networks. This network decays the slowest. Moreover its  $p$-core volume stays unchanged at $83.2\pm 1.3 \% $ for  $ 8 \% < p < 18\%$. This firmed core is built from transitions between the three vertices $0$ and $\pm 8$. A similar network core also emerges from the network constructed from surrogate signals of the $HTX$ group. Its volume of $76.8 \pm 1.5 \% $ is significantly lower than the volume of the network constructed from original cardiac signals (by Mann-Whitney $U$ test, $P=0.011$) for all $p$ in the interval described. This fact may indicate that the network organization resulting from the cardiac signals relies on important nonlinear dependences.

On the contrary a similar property does not hold in the case of the disintegration of the network constructed from the {\it Young} group signals. The volume of the $p$-core decays in the same way in both networks: the network made from cardiac signals and the network made from surrogate signals. The disintegration of both networks goes fast, for example at $p=10\%$, the network volume is about $5\pm 2\%$ in both cases.

Obviously, the networks obtained from shuffled signals decay more quickly when compared to the networks produced from cardiac series. 

\subsection{Adjacency and transition matrices for the {\it Young} }
\begin{figure}[htbp]
\centering
  \includegraphics[width=0.95\textwidth ]{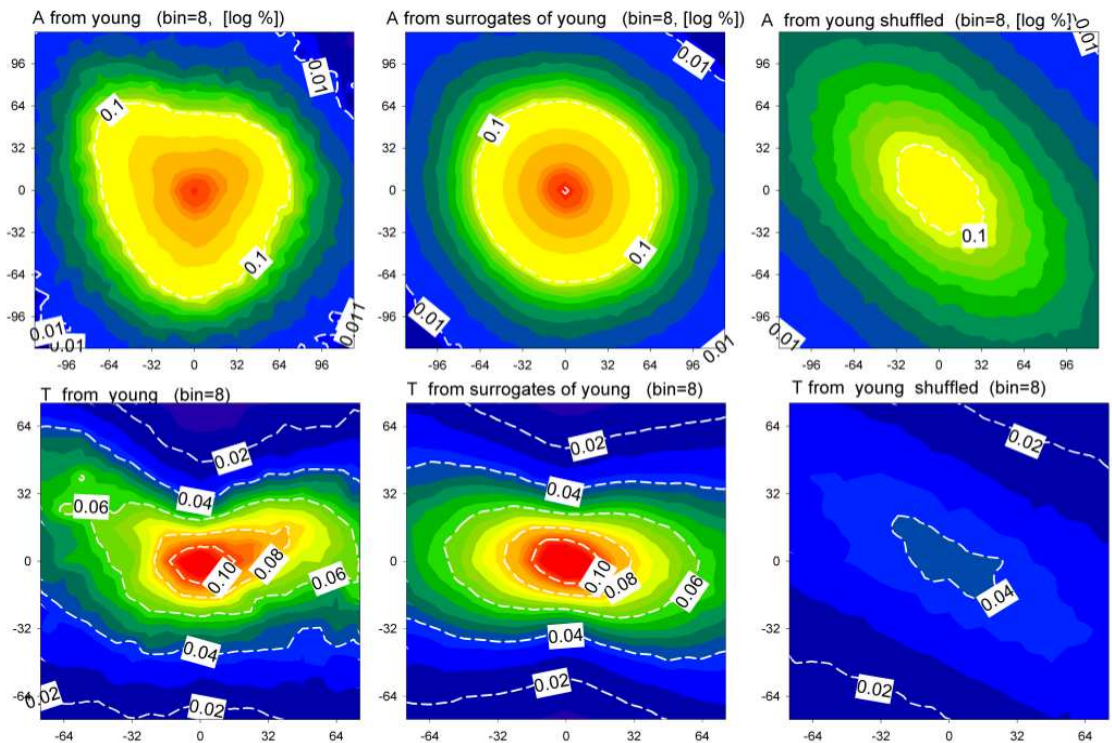}
    \caption{Adjacency matrices $\bf A$  (first row) and transition matrices $\bf T$ (second row) for cardiac signals of the {\it Young} group (left column),  for their {\it surrogates} (middle column)  and {\it shuffled} signals (right column). Contour representations  of $\bf A$  show  logarithms of  $\bf A$ values given in per cents from 0.001\% to 1\%. Contours representing  $\bf T$ are plotted in linear scale from 0.0 to 0.1 with step 0.02.}
    \label{fig:adjacency}
 \end{figure}

In Fig. \ref{fig:adjacency}, we show contour plots of both matrices: adjacency $\bf A$ and transition $\bf T $  obtained from  signals recorded from the {\it Young} group. Together, the plots obtained from surrogates of these  signals and shuffled $RR$-intervals are presented.  All the plots represent signals binned at $\Delta_{0}$ and for changes smaller than 100 ms for $\bf A$ and smaller than 70 ms in case of $\bf T$.

From the plots in Fig. \ref{fig:adjacency}, it is evident that shuffled $RR$-intervals provide different matrix representations.  The meaningful smaller probability of events corresponding to small accelerations or decelerations (i.e., probability of events around $(0,0)$) is a noticeable effect of the independence of $RR$-intervals.  Moreover, the symmetry in these matrices is different from symmetries which are present in other plots. These symmetric features can be explained by elementary counting of the sets built from the three $RR$-interval events of the specific type.  Since the state space of $RR$-intervals is discrete and limited, the values of $RR_{t} $ for any $t=1,2,\dots, N$ take one of the values from the set of $K'$ different values:
\begin{equation}
RR^{(1)}= \min {\bf RR_{\Delta_{bin}}} < RR^{(2)}= RR^{(1)} + \Delta_{bin} < \quad \dots  < \quad RR^{(K')}= \max {\bf RR_{\Delta_{bin}}}.
\label{min_max}
\end{equation}
For simplicity, let us assume that all events of Eq. \ref{min_max} are equally probable, $p (RR^{(I)}) = K'\slash N $. 
Then the number of possible monotonically growing three-element sequences ($RR^{(I)}$, $RR^{(J)}$, $RR^{(M)}$) with $I < J < M$ constructed from these values, namely patterns of  the  2LV type,   is $ K'(K'-1)(K'-2)\slash 3$. On the other hand, from each sequence of the 2LV type, one can construct two different three-element sequences of the 2UV type. Hence the total probability of events of the 2UV type is twice as large as of the 2LV type. 

In the case when events from the list described in Eq. \ref{min_max} are not distributed uniformly, the calculations are more demanding but finally they lead to the conclusion that if $\Delta^{(I)}\Delta^{(J)} < 0$, then events  $(\Delta^{(I)}, \Delta^{(J)})$ are more probable than  $(-\Delta^{(I)},\Delta^{(J)} )$ or $(\Delta^{(I)},-\Delta^{(J)} )$.

In addition, comparing $\bf A$ matrices obtained from cardiac signals and their  surrogates, Fig. \ref{fig:adjacency} first row,  we see:
\begin{itemize}
\item there is a small deficiency  of events close to $(0,0)$ in cardiac signals when compared to signals with  surrogates;
\item there are three regions in $\bf A$ in which cardiac signals dominate their surrogates. They can be described as three $RR$-interval events of the type:
\begin{itemize}
 \item[(1)]  2LV  but related only to  large decelerations;
 \item[(2)]  1V   when after a small change, a large acceleration occurs;
 \item[(3)]  2UV  but only for a large deceleration after a large acceleration.
\end{itemize}
 By large changes above, we mean $RR$-increments greater than 30 ms.
\end{itemize}

\subsection{Adjacency and transition matrices for {\it HTX} }
\begin{figure}[htbp]
\centering
 \includegraphics[width=0.95\textwidth ]{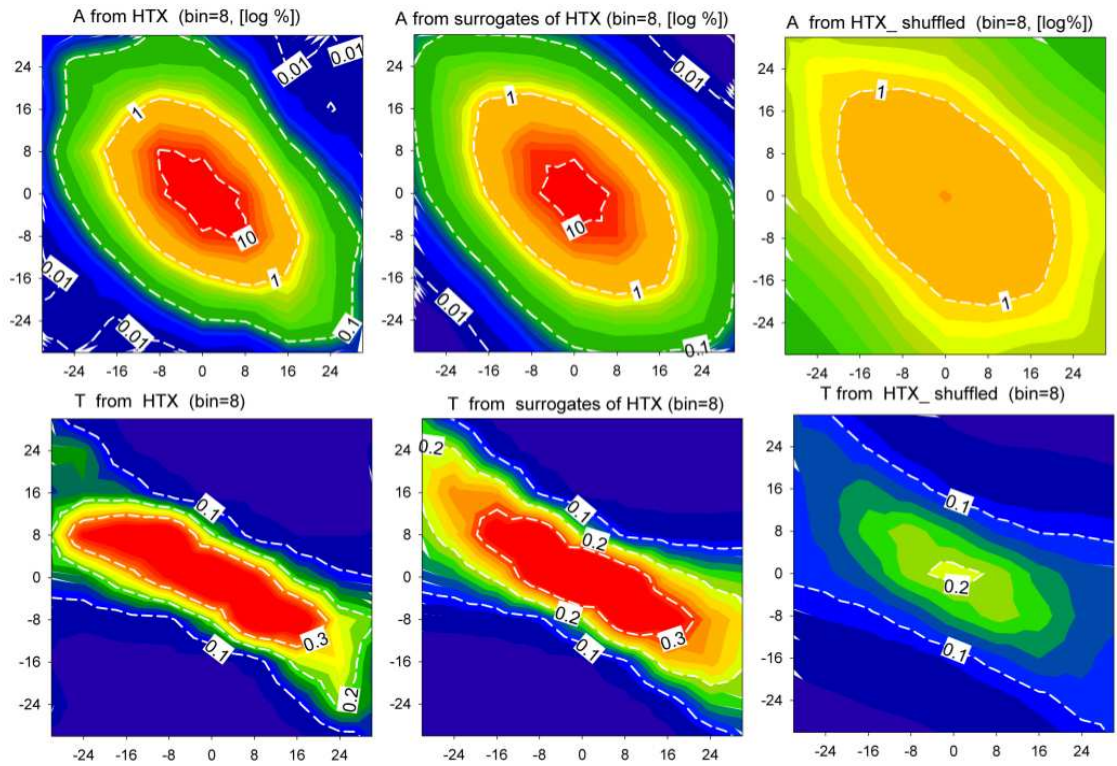}
\caption{Adjacency matrices $\bf A$ (first row) and transition matrices $\bf T$ (second row) for cardiac signals of  {\it HTX} group (left column), for their {\it surrogates} (middle column) and {\it shuffled} signals (right column). Values of $\bf A$ are shown as logarithms of per cents from 0.01\% to 10\%.   $\bf T$s are plotted in a linear scale from 0.0 to 0.3 with step 0.1.}
\label{fig:adjacency_htx}
\end{figure}

In the absence of any influence of the autonomic nervous system, the network representation of $RR$-increments consists of considerably fewer vertices than for a typical healthy person, as has been already shown in Fig. \ref{fig:htx}.
Fig. \ref{fig:adjacency_htx} presents results aimed at widening our understanding of the  nonlinear effects of the intrinsic mechanisms controlling  the  heart contractions. Note that the contour plots in Fig. \ref{fig:adjacency_htx} are in different scales from the plots representing signals of the {\it Young} group in Fig. \ref{fig:adjacency}.

We see in Fig. \ref{fig:adjacency_htx}  similar symmetric features in all three plots with $\bf A$ matrices, namely that alternating changes with  $ \Delta^{(I)} \cdot \Delta^{(J)} < 0$ are more dominant than monotonic changes. Following our discussion in the previous subsection about the imprints of randomness, this observation may imply stochastic independence of the underlying dynamics. However, the cardiac dynamics is more concentrated around transitions from a no-change event to the smallest increments possible, namely to $\pm 8, \pm 16$ ms, than if it resulted from random sources. Furthermore, a closer analysis of $\bf A$ (compare Figs \ref{fig:htx}(A) and \ref{fig:htx}(B)) reveals that the system represented by the cardiac signals is less likely to stay in the no-change vertex, and changes of size $16$ ms between vertices representing transitions of $ +8$ and  $- 8 $ ms in size occur more often in the cardiac signals [$p(8,-8)=p(-8,8) =11\pm 1\%$] than in their  surrogate series [$p(8,-8)\approx p(-8,8)\approx 8.4\pm 0.2\%$]. 

Additionally, a comparison between the corresponding  $\bf T$ matrices provides important distinctions between cardiac signals and their surrogates in the system reaction after the larger accelerations, namely, if $\Delta < -16 $ ms. It appears that when accelerating, the system is more resistant to a pendulum-like reaction. This has the effect that the $RR$-interval is able to retain the shorter rhythm for the next contraction.

\begin{figure}[htbp]
\centering
 \includegraphics[width=0.9\textwidth ]{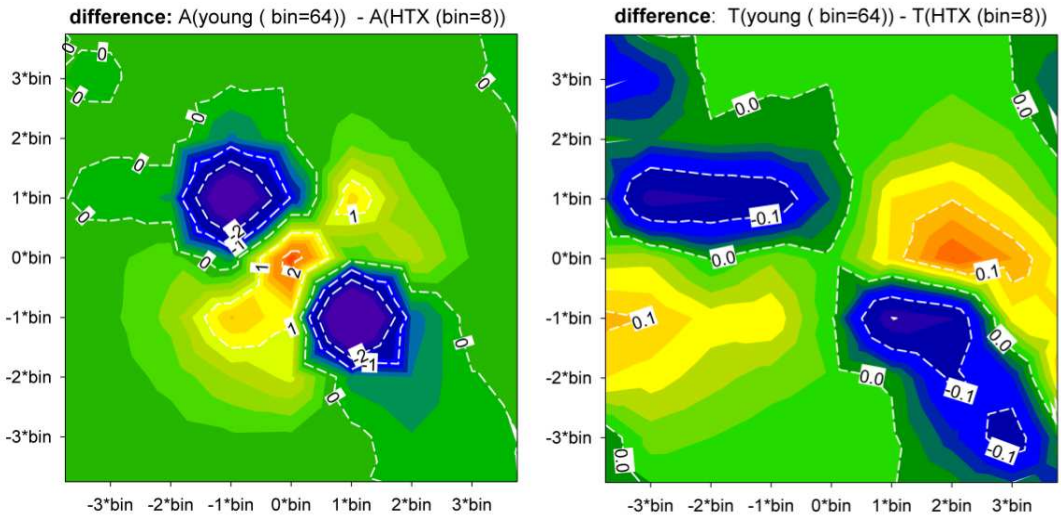}
  \caption{Difference between adjacency matrices (left) and transition matrices (right) obtained from cardiac signals of the {\it Young} group when the signals were binned in $\Delta_{bin}=64$ ms  and from recordings of the {\it HTX} group. Values of differences in $\bf A $ are in per cents. The more reddish areas are the regions of greater cardiac superiority. The more blue areas are the regions of greater superiority of surrogates.}
\label{fig:difference}
 \end{figure}

In Fig. \ref{fig:difference}, we show plots of differences between matrix graphs arising from cardiac signals of the {\it Young} group when the signals are binned in $\Delta_{bin}=64$ ms, and the {\it HTX} group. We see that a similarity is apparent between the dynamics underlying these two systems. The basic distinction relies on events of the 2LV type, where two subsequent accelerations or two subsequent decelerations are involved. Thus it can be hypothesized that independently of the scale, the purpose of the autonomic regulation seems to be to speed up or slow down the heart period. Since such persistence could be involved in some overall purpose like actual bodily needs, the next supposition can be formulated as follows.  While the intrinsic heart control mechanisms are devoted to keeping the homeostasis, the control of the autonomic nervous system aims at satisfying physical demands.

\section{Conclusions}
Network structure methods are able to visualize, describe and differentiate heart rate dynamics in healthy young subjects and HTX patients. Using these methods, the general dynamical properties of heart rate can be defined as correlated or not correlated (employing the comparison of raw and shuffled signals) and linearly or non-linearly correlated (comparing raw signals and signals with shuffled phases of Fourier Transform).

The essential feature of complex dependencies in  nocturnal heart rhythm in our group of healthy young persons is related to large $RR$-increments, both decelerations and accelerations. This feature manifests itself in that large accelerations are more likely antipersistent, while large decelerations are more likely persistent. This observation also seems to be an important indicator of healthy heart rate. 

Moreover, since the vagal part of autonomic regulation is considered responsible for large $RR$-increments, we may hypothesize that vagal activity is a crucial source of complexity in short-term heart rate variability. In healthy young individuals, the change in vagal tone during sleep (e.g. change from high vagal activity to its withdrawal between non-REM and REM sleep stages) allows us to observe the specific patterns of heart rate dynamics. We interpret the non-linear relationship observed between consecutive accelerations and decelerations in the case of bigger changes (accelerations and decelerations of more than 35 ms) as an effect of vagal activity.

Although in HTX patients' heart rate regulation is mostly intrinsic with no autonomic control, the relationship between consecutive accelerations and decelerations is also observed, but in this case the scale of changes is much lower. $RR$-increments vary as fluctuations around a homeostatic state. However, the organization of this homeostatic state in the case of raw signals shows that it involves dynamical forces more strongly than if the dynamics were driven by linear forces only. In post-transplant patients, the non-linear dependencies are also characterized by the appearance of sequences made of bigger ($>20$ ms) accelerations followed by smaller decelerations (less than 10 ms). This means that an increase in heart rate is not so effective as in healthy individuals but is still possible. We hypothesize that this pattern of heart rate in HTX patients may be a result of gradual sympathetic reinnervation.

\end{document}